\documentclass[showpacs,preprintnumbers,amsmath,aps,amssymb]{revtex4}
\usepackage{graphicx}
\usepackage{dcolumn}
\usepackage[english]{babel}
\usepackage{bm}
\begin{document}
\title{FRW in cosmological self-creation theory }
\author{Juan M. Ram\'irez$^1$}
\email{jmramirz@fisica.ugto.mx}
\author{J. Socorro$^{1,2}$}
\email{socorro@fisica.ugto.mx} \affiliation{$^1$Departamento de
F\'{\i}sica, DCeI, Universidad de Guanajuato-Campus Le\'on,
 C.P. 37150, Le\'on, Guanajuato, M\'exico\\
  $^2$ Departamento de F\'isica, Universidad Aut\'onoma Metropolitana, Apartado Postal 55-534,
C.P. 09340 M\'exico, DF, M\'exico
}%

\date{\today}

\begin{abstract}
We use the Brans-Dicke theory from the framework of General Relativity (Einstein frame),
but now the total energy momentum tensor fulfills the following condition
$\rm \left[\frac{1}{\phi}T^{\mu \nu M}+T^{\mu \nu}(\phi)\right]_{;\nu}=0$.
We take as a first model the flat FRW metric and with the law of variation
for Hubble's parameter proposal by Berman \cite{Berman}, we find solutions
to the Einstein field  equations by the cases: inflation ($\gamma=-1$),
radiation ($\gamma=\frac{1}{3}$), stiff matter ($\gamma=1$). For the Inflation
case the scalar field grows fast and depends strongly of the constant
$\rm M_{\gamma=-1}$ that appears in the solution, for the Radiation case,
the scalar stop its expansion and then decrease perhaps due to the
presence of the first particles. In the Stiff Matter
case, the scalar field is decreasing so for a large time, $\phi\rightarrow0$. In the same line of classical solutions,
we find an exact solution to the Einstein field equations for the stiff matter
$(\gamma=1)$ and flat universe, using the Hamilton-Jacobi scheme.
\end{abstract}

\maketitle
\title{}
\maketitle
\section{Introduction}
The Scalar-Tensor theories have their origin in the $50's$. Pascual
Jordan was intrigued by the appearance of a new scalar field in
Kaluza-Klein theories, especially in its possible role as a
generalized gravitational constant. In this way appear the
Brans-Dicke theory, with the particularity that each one energy
momentum tensor satisfy the covariant derivative \cite{Weinberg},
$\rm T^{\mu \nu i}\,_{;\nu}=0$, where $i$ corresponds to the i-th
ingredient of matter content. Late years ago (1982), appear a new
proposal by Barber \cite{Barber1}, known as self-creation cosmology
(SCC) \cite{Barber,SSC10}. Since
 the original paper  appeared in 1982, more and more authors \cite{Singh,Singh0,Singh1,Singh2} have worked the different versions of this theory in the
 classical fashion. By instant, Singh and Singh \cite{Singh3} have studied Raychaudhary-type
equations for perfect fluid in self-creation theory. Pimentel
\cite{Pimentel} and Soleng \cite{Soleng1,Soleng2} have studied in
detail the cosmological solutions of Barber's self-creation
theories. Reddy \cite{Reddy}, Venkateswarlu and Reddy \cite{Venka1},
Shri and Singh \cite{Shri1,Shri2}, Mohanty et al. \cite{Mohanty},
Pradhan and Vishwakarma \cite{Pradhan1,Pradhan2}, Sahu and Panigrahi
\cite{Sahu}, Venkateswarlu and Kumar \cite{Venka2} are some of the
authors who have studied various aspects of different cosmological
models in self-creation theory. These papers adapted the Brans Dicke
theory to create mass out of the universe's self contained scalar,
gravitational and matter fields in simplest way.

 The
gravitational theory must be a metric theory, because this is the
easiest way to introduce the Equivalence Principle. However always
is possible to put additional terms to the metric tensor, the most
obvious proposal is a scalar field $\phi$. Recently Chirde and
Rahate \cite{Rahate} investigated spatially homogeneous isotropic
Friedman-Robertson-Walker cosmological model with bulk viscosity and
zero-mass scalar field in the framework of Barber's second
self-creation theory and found classical solutions, because is the
simplest way to work this theory, because only take in account the
energy-momentum tensor of usual matter and the scalar $\phi$.

This work is arranged as follow. In section II we present the method
used in general way, where we take the Brans-Dicke Lagrangian
density and consider this theory in self-creation theory. In section
III, employing the flat FRW cosmological model as a toy model with
barotropic perfect fluid and cosmological constant and found
classical solutions for three epoch in our universe, inflation
phenomenon $(\gamma=-1)$, radiation $(\gamma=\frac{1}{3})$ and stiff matter
$(\gamma=1)$. In Section IV we construct the Lagrangian and
Hamiltonian densities for the cosmological model under consideration
and are presented other class of classical solutions using the
Hamilton-Jacobi approach. The section V is devoted to the conclusions
of the work.
\section{Self Creation Cosmology in GR}
The Lagrangian density in the Brans-Dicke theory is
\begin{equation}
{\cal L}[g,\phi]= \frac{\sqrt{-g}}{16\pi}\left(R\phi -\frac{\omega}{\phi}g^{\mu\nu}\phi_{,\mu}\phi_{,\nu}\right)+\sqrt{-g}L_{matter},\label{L1}
\end{equation}
where $L_{matter}=16\pi \rho$, so making the corresponding variation to the scalar field and the tensor metric, the field equations in this theory become
\begin{equation}
R-\frac{\omega}{\phi^2} g^{\mu\nu}\phi_{,\nu}\phi_{,\mu}+\frac{2\omega}{\phi}\square^2\phi=0,\label{EC1}\\
\end{equation}

\begin{equation}
\rm
R^{\alpha\beta}-\frac{1}{2}g^{\alpha\beta}R=\frac{8\pi}{\phi}T^{\alpha\beta}
+\frac{\omega}{\phi^2}\left(\phi^{,\alpha}\phi^{,\beta}-\frac{1}{2}g^{\alpha\beta}\phi^{,\lambda}\phi_{,\lambda}\right)
+\frac{1}{\phi}\left(\phi^{,\alpha;\beta} -
g^{\alpha\beta}\square^{2}\phi \right),\label{ECG}
\end{equation}
where the left side is  the Einstein tensor, the first term in the right side is the corresponding energy momentum tensor of material mass coupled
with the scalar field $\phi^{-1}$. The second and third term corresponds at energy momentum tensor of the scalar field coupled also to $\phi^{-1}$.
 Both equations are recombined given the relation
 \begin{equation}
 \rm \Box \phi=4\pi \lambda T, \label{dalamber}
 \end{equation}
 where $\lambda$ is a coupling constant to be determined from experiments.
The measurements of the deflection of light restrict the value of
coupling to $|\lambda| < 10^{-1}$. In the limit $\lambda \to 0$, the
Barber's second theory approaches the standard general relativity
theory in every respect. $\Box \phi = \phi_{;\mu}^{;\mu}$ is the
invariant D'Alembertian and T is the trace of the energy momentum
tensor that describes all non gravitational and non scalar field
matter and energy. Taking the trace of the equation (\ref{ECG}) and
then substitute in (\ref{dalamber}), we obtain the following wave
equation to $\phi$
\begin{equation}\label{e:SSC1}
\Box\phi=\frac{8\pi}{(3+2\omega)}T_{M},
\end{equation}
comparing both equations (\ref{dalamber}) and (\ref{e:SSC1}), we note that $\lambda=\frac{2}{(3+2\omega)}$, where $\omega$ is a coupling constant.\\
Now the Einstein equation can be rewritten as
\begin{equation}
\rm G^{\alpha \beta}=\frac{8\pi}{\phi}T^{\alpha\beta (M)} +\frac{1}{\phi} T^{\alpha \beta (\phi)}=T^{\alpha \beta (T)},
\end{equation}
where
$$\rm T^{\alpha \beta (\phi)}=\frac{\omega}{\phi}\left(\phi^{,\alpha}\phi^{,\beta}-\frac{1}{2}\phi^{,\lambda}\phi_{,\lambda}g^{\alpha\beta}\right)
+\left(\phi^{,\alpha;\beta} - g^{\alpha\beta}\square^{2}\phi
\right).
$$
In Brans-Dicke theory, each one energy momentum tensor satisfy the
covariant derivative, $\rm T^{\mu \nu}\,_{;\nu}=0$.
 In the self creation theory, we introduce that the total energy momentum tensor is which satisfy the covariant
 derivative, $\rm T^{\mu \nu (T)} \,_{;\nu}=0$, namely;
\begin{equation}\label{e:tensor}
\rm T^{\alpha \beta (T)} \,_{;\beta}=0, \qquad \Rightarrow \qquad
 \left[\frac{8\pi}{\phi}T^{\alpha\beta (M)} +\frac{1}{\phi} T^{\alpha \beta (\phi)}\right]_{;\beta}=0,
 \end{equation}
which imply that $\rm \left[ Q^{\alpha \beta (T)}
\right]_{;\beta}=\frac{\nabla_\beta \phi}{\phi} \left[ Q^{\alpha
\beta (T)} \right]$,  where $\rm Q^{\alpha \beta (T)}= \phi
T^{\alpha \beta (T)}$. This equation is the master equation which
gives the name of  self creation theory, because the covariant
derivative of this tensor have a source of the same tensor multiply
by a function of the scalar field $\rm \phi$.

\section{FRW in Self Creation theory}
We apply the formalism using the geometry of Friedmann-Robertson-Walker
\begin{equation}
 \rm  ds^2  =  - N^2(t) dt^2  + A^2(t)  \left[\frac{dr^2}{1-\kappa r^2} + r^2 d\theta ^2  + r^2 sen^2 (\theta )d\varphi ^2\right],
 \label{metric}
\end{equation}
where N is the lapse function, A is the scalar factor. Now we solve the equations (\ref{ECG}), (\ref{dalamber}) and (\ref{e:tensor}),
with the aim to find solutions to Density $\rho(t)$, Scalar factor $A(t)$ and the Scalar field $\phi(t)$.\\
Taking the transformation $\prime=\frac{d}{Ndt}=\frac{d}{d\tau}$ and using $T^{\alpha\beta}$ as a fluid perfect,
first compute the classical field equations (\ref{ECG}), together with the barotropic equation of state $P=\gamma \rho$, this equation become
\begin{equation}\label{e:ec3}
3\left(\frac{A'}{A}\right)^2+3\frac{A'}{A}\frac{\phi'}{\phi}-\frac{1}{2}\omega\left(\frac{\phi'}{\phi}\right)^2
-8\pi\frac{\rho}{\phi}+3\frac{\kappa}{A^2}=0,
\end{equation}
\begin{equation}\label{e:ec4}
2\frac{A''}{A}+\frac{A'^2}{A^2}+2\frac{A'}{A}\frac{\phi'}{\phi}+\frac{\phi''}{\phi}+\frac{1}{2}\omega\frac{\phi'^2}{\phi^2}
+8\pi\gamma\frac{\rho}{\phi}+\frac{\kappa}{A^2}=0,
\end{equation}
the equation (\ref{dalamber}) become
\begin{equation}
 3\frac{A^\prime}{A} \frac{\phi^\prime}{\phi}+ \frac{\phi^{\prime\prime}}{\phi} = 4 \pi \lambda (1-3\gamma)\frac{\rho}{\phi}.\label{dal}
\end{equation}
The covariant equation (\ref{e:tensor}) or conservation equation to
the total energy momentum take the following form
\begin{equation}
\label{e:econserv} \rm
-3\frac{A''}{A}\frac{\phi'}{\phi}+3\omega\frac{A'}{A}\left(\frac{\phi'}{\phi}\right)^2
-\omega\left(\frac{\phi'}{\phi}\right)^3+3\frac{A'}{A}\left(\frac{\phi'}{\phi}\right)^2
+24\pi\frac{A'}{A}(1+\gamma)\frac{\rho}{\phi}+\omega\frac{\phi''}{\phi}\frac{\phi'}{\phi}-
8\pi\frac{\phi'}{\phi}\frac{\rho}{\phi}+8\pi\frac{\rho'}{\phi} = 0,
\end{equation}
thus, the system equations to be solved are (\ref{e:ec3})-(\ref{e:econserv}).\\
We write the term $\left(\frac{ A^\prime}{A}
\right)^2+\frac{\kappa}{A^2}$, in equation (\ref{e:ec3}), using
equations (\ref{e:ec4}) and (\ref{dal}), and after some algebra we
have
\begin{equation}
\rm 3\frac{A^{\prime\prime}}{A}-3\frac{A^{\prime}}{A} \frac{\phi^\prime}{\phi}+\omega\left(\frac{ \phi^\prime}{\phi} \right)^2 =
 -4\pi\left[\frac{3}{2}\lambda(1-3\gamma)+(1+3\gamma)\right]\frac{\rho}{\phi}. \label{equ}
 \end{equation}
Equation (\ref{e:econserv}) can be rewritten as
\begin{equation}
\label{e:covar}
\rm\left(-3\frac{A''}{A}+3\frac{A'}{A}\frac{\phi'}{\phi}-\omega\left(\frac{\phi'}{\phi}\right)^2\right)
\frac{\phi'}{\phi}+\left(3\frac{A'}{A}\frac{\phi'}{\phi}+\frac{\phi''}{\phi}\right)\frac{\phi'}{\phi}\omega
+24\pi(1+\gamma)\frac{\rho}{\phi}\frac{A'}{A}+8\pi\left(\frac{\rho}{\phi}\right)'=0,
\end{equation}
  and using the equations (\ref{dal}) and (\ref{equ}) in
(\ref{e:covar}), then we have the master equation for solve the
energy density of the model, as
\begin{equation}
\rm
2\pi\left[2(1+3\gamma)+\lambda(1-3\gamma)(3+2\omega)\right]\frac{\rho}{\phi}\frac{\phi'}{\phi}
+24\pi(1+\gamma)\frac{\rho}{\phi}\frac{A'}{A}+8\pi\left(\frac{\rho}{\phi}\right)'=0,
\end{equation}
 defining the function $\rm F=\frac{\rho}{\phi}$, we have
 \begin{equation}
 \rm \frac{d}{d\tau} Ln \left[F A^{3(1+\gamma)}
 \phi^{\frac{1}{4}[2(1+3\gamma)+\lambda(1-3\gamma)(3+2\omega)]}\right]=0,
 \end{equation}
  who solution is
\begin{equation}
 \rm \rho=M_\gamma  A^{-3(1+\gamma)} \phi^\beta, \qquad \beta=\frac{(1-3\gamma)}{4}\left[2-\lambda a_0
 \right], \qquad a_0=3+2\omega.
\label{rho}
\end{equation}
This equation is equivalent to General Relativity expression
\cite{Barber} with the addition the last factor representing
 the self creation cosmology. We note that $\rm \lambda_{0}$ is a new free parameter, however if $\rm \lambda_{0}=\lambda=\frac{2}{3+2\omega}$,
 then $\rm \beta=0$ so equation (\ref{rho}) became the usual solution to General Relativity. On the other hand for a photon gas $\gamma=\frac{1}{3}$, we have $\beta=0$
 and equation (\ref{rho}) reduce to its General Relativity expression $\rm \rho=\rho_{0}(A/A_{0})^{-4}$ which is consistent,
 because in the radiation epoch there was not interaction between photon and the scalar
 field.

\begin{figure}
\begin {center}
\includegraphics[width=8.5cm]{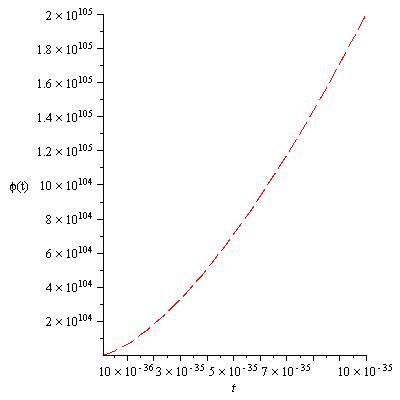}
\caption{The shape of Scalar field is the same for $0<n<1$, the value of the constant $M_{-1}$
tells us how the scalar field grows, curve reaches the highest and lowest altitude for large and small
values of the constant $M_{-1}$ respectively. In this case $n=0.8$ and $M_{-1}=10^{75}$}\label{inf-scalar}
\end {center}
\end{figure}

 Now we focus on finding the scalar field and scale factor $\phi(t)$, $A(t)$.
Berman and Gomide \cite{Berman} proposed the following law of variation for Hubble's parameter
\begin{equation}\label{e:hubble}
\rm H=\frac{\dot{A}}{A}=DA^{-n},
\end{equation}
where D and n are constants, $\rm \dot{}=\frac{d}{dt}$, using the
following definition
\begin{equation}\label{e:desace} \rm
q=-\frac{\ddot{A}A}{\dot{A}^2},
\end{equation}
where q is the deceleration parameter.
\\Using (\ref{e:hubble}) into (\ref{e:desace}) we have
\begin{equation}\label{e:desace1}
\rm q=n-1, \qquad n=q+1,
\end{equation}
the relation (\ref{e:desace1}) imply that q as a constant. The sign
of q indicated whether the model inflates or not. The positive sign
of q i.e. $\rm q>0$ correspond to standard decelerating model,
whereas the negative sign $\rm -1\leq q < 0$ for $\rm 0\leq n < 1$
indicates inflation \cite{chohuan}. Many authors (see, Singh et al.
\cite{Singh,Singh1,Singh2}), have studied flat FRW and Bianchi type
models by using the special law for Hubble parameter that yields
constant value of deceleration parameter. The equation
(\ref{e:hubble}) writes as
\begin{equation}\label{e:escalafac} \rm
\dot{A}A^{n-1}=D.
\end{equation}
Then solving the equation (\ref{e:escalafac}) we obtain the law for average scale factor $A(t)$ as
\begin{equation}\label{e:escala1} \rm
  A(t)=C_{1}e^{Dt}, \qquad n=0, \\
\end{equation}
\begin{equation}\label{e:escala2} \rm
  A(t)=[nDt+C_{2}]^{1/n}, \qquad n\neq0,
\end{equation}
where $\rm C_{1}$ and $\rm C_{2}$ are constants of integration.
Equation (\ref{e:escala2}) implies
that the condition for the expansion of the universe is $\rm n = q + 1 > 0$.\\
With the equation (\ref{rho}) and the Berman's law (\ref{e:escala2})
we can find the solution to scalar field $\rm \phi(t)$, we can use
$C_{2}=0$ in our next calculations, we may recover the general
results by the substitution $\rm nDt=C_{2}'+nDt'$. Taking account
again the classical field equations with $\rm \Lambda=0$ and
$\kappa=0$, adding equation (\ref{e:ec3}) and (\ref{e:ec4}) with the
gauge $N=1$ we get
\begin{equation}\label{e:escala3} \rm
4\frac{\ddot{A}}{A}+8\left(\frac{\dot{A}}{A}\right)^2
+10\frac{\dot{A}}{A}\frac{\dot{\phi}}{\phi}+2\frac{\ddot{\phi}}{\phi}+16\pi(\gamma-1)\frac{\rho}{\phi}=0,
\end{equation}
so let's solve the equation (\ref{e:escala3}) for the following cases:\\
\\ \textbf{Case I}\\
Inflation ($\rm \gamma=-1$), then the equation (\ref{e:escala3}) is
written as
\begin{equation}\label{e:escala5}
\rm
2\frac{\ddot{A}}{A}+4\left(\frac{\dot{A}}{A}\right)^2+5\frac{\dot{A}}{A}\frac{\dot{\phi}}{\phi}
+\frac{\ddot{\phi}}{\phi}-16\pi\frac{\rho}{\phi}=0,
\end{equation}
now inserting (\ref{e:escala2}) and (\ref{rho}) into
(\ref{e:escala5}), after some algebra we have
\begin{eqnarray}
\rm \nonumber t^2\ddot{\phi}+\frac{5}{n}t\dot{\phi}+\left[\frac{6-2n}{n^2}\right]\phi=16\pi M_{-1}t^2\phi^{\beta},\\
\rm t^2\ddot{\phi}+at\dot{\phi}+b\phi=ct^2\phi^{\beta},\label{e:CasoII1}
\end{eqnarray}
where $\rm \beta=2-\lambda_{0}(3+2\omega)$, the equation
(\ref{e:CasoII1}) has the form $\rm x^2y''+axy'+by=cx^ny^m$. We
solve this equation for the case $\rm \beta=1$, so the equation
(\ref{e:CasoII1})  is rewritten as
\begin{eqnarray}\label{e:CasoII2}
\nonumber \rm t^2\ddot{\phi}+\frac{5}{n}t\dot{\phi}+\left[\frac{6-2n}{n^2}\right]\phi-16\pi M_{-1}t^2\phi=0,\\
\nonumber  \rm t^2\ddot{\phi}+at\dot{\phi}+(ct^2+b)\phi=0,
\end{eqnarray}
which solution is \cite{polyanin}
\begin{equation}
\rm
\phi(t)=t^{\frac{1-a}{2}}[C_{1}I_{\nu}(\sqrt{c}t)+C_{2}K_{\nu}(\sqrt{c}t)],\nonumber
\end{equation}
where$\rm I_{\nu}$, $\rm K_{\nu}$
 are a modified Bessel functions of first and second kind respectively, and $\rm a=\frac{5}{n}$, $\rm b=\frac{6-2n}{n^2}$, $\rm c=16\pi M_{-1}$,
  $\rm \nu=\frac{1}{2}\sqrt{(1-a)^2-4b}$. Now we are interesting in the inflation time $\rm (10^{-36}s\sim10^{-34}s)$,
 $\rm K_{\nu}$ diverges near the origin, then $\rm C_{2}=0$, so the solution to (\ref{e:CasoII2}) is
\begin{equation}
\rm
\phi(t)=t^{\frac{1-a}{2}}[C_{1}I_{\nu}(\sqrt{c}t)].\label{e:escinfla}
\end{equation}
Figure (\ref{inf-scalar}) shows the shape of the scalar field $\phi$ in the inflation time. The behavior of the scalar field
is the same for $0<n<1$ and depends strongly of the constant $M_{\gamma=-1}$ that appear in the solution (\ref{rho})
and (\ref{e:escinfla}).
\\This solution was found with $\beta=1$, so in this case the physical quantities take a form
\begin{equation}\rm
-P=\rho=C_{1}M_{-1} t^{\frac{(1-a)}{2}}[I_{\nu}(\sqrt{c}t)].
\end{equation} \\

\begin{figure} [h]
\begin {center}
\includegraphics[width=8.5cm]{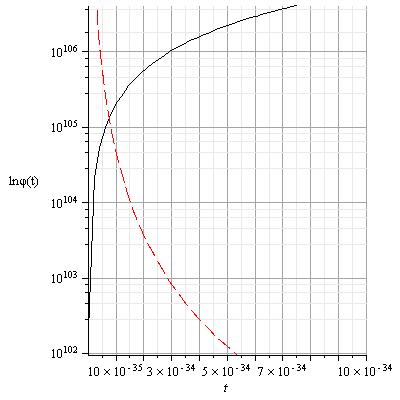}
\caption{$\rm Log(\phi)vs\,t$.  The scalar field is increasing in
the inflation epoch, but in the radiation time the scalar field
decrease.}\label{infla-rad}
\end {center}
\end{figure}

\textbf{Case II} \\
Radiation $\rm (\gamma=\frac{1}{3})$. In this case the equation
(\ref{e:escala3}) is written as
\begin{equation}\rm
2\frac{\ddot{A}}{A}+4\left(\frac{\dot{A}}{A}\right)^2+5\frac{\dot{A}}{A}\frac{\dot{\phi}}{\phi}
+\frac{\ddot{\phi}}{\phi}-\frac{16}{3}\pi\frac{\rho}{\phi}=0,\label{e:escala6}
\end{equation}
now (\ref{e:escala2}) and $\frac{\rho}{\phi}$ into
(\ref{e:escala6}), after some algebra we have
\begin{eqnarray}
\rm \nonumber t^{2}\ddot{\phi}+\frac{5}{n}t\dot{\phi}+\left[\frac{6-2n}{n^2}\right]\phi=\frac{16}{3}\pi \frac{M_{1/3}}{(n D)^{4/n}}t^{2-\frac{4}{n}}, \\
\rm t^2\ddot{\phi}+at\dot{\phi}+b\phi=dt^m,  \label{e:radiation}
\end{eqnarray}
where $\rm a=\frac{5}{n}$, $\rm b=\frac{6-2n}{n^2}$, $\rm d=\frac{16\pi}{3}\frac{M_{1/3}}{[nD]^{4/n}}$, and $\rm m=\frac{2n-4}{n}$.\\
We need to solve the equation (\ref{e:radiation}), so taking account
the following transformation $\rm t=e^x$, then (\ref{e:radiation})
is rewritten as
\begin{equation}
\rm \frac{d^2\phi}{dx^2}+a_{1}\frac{d\phi}{dx}+b\phi=de^{mx},\label{e:transf1}
\end{equation}
where $\rm a_{1}=a-1$, now using the next transformation
\begin{equation}\rm
\phi=\omega e^{-\frac{1}{2}a_{1}x}.\label{e:trans-phi}
\end{equation}
So the equation (\ref{e:transf1}) is now
\begin{equation}
\rm \frac{d^2\omega}{dx^2}+f\omega=de^{mx+\frac{1}{2}a_{1}x}, \qquad f=\left(b-\frac{1}{4}a_{1}^{2}\right).\label{e:transf2}
\end{equation}
The solution of equation (\ref{e:transf2}) depend the sing of the constant $f$, but remembering that $0<n<1$, so $f=-\left(\frac{n-1}{2n}\right)^2<0$, then the solution is
\begin{eqnarray}\rm
\nonumber \omega=C_{1}\cosh(k_{1}x)+C_{2}\sinh(k_{1}x)\\
\rm
+\frac{d}{k_{1}}\int_{x_{0}}^{x}e^{(m+\frac{a_{1}}{2})\xi}\sinh[k_{1}(x-\xi)]d\xi,
\label{e:sol-a}
\end{eqnarray}
where $\rm x_{0}$ is an arbitrary number and $\rm f=-k_{1}^2<0$. Now
solving the integral that appear in  equation (\ref{e:sol-a}) with
$\rm \sigma=m+\frac{a_{1}}{2}$, we have the solution to $\omega$ is
the following
\begin{eqnarray}\rm
\nonumber \omega(x)=C_{1}\cosh(k_{1} x)+C_{2}\sinh(k_{1} x)+\\
\rm \frac{d}{k_{1}}\left[\frac{k_{1}}{\sigma^2-k_{1}^2}e^{\sigma
x}+\frac{e^{(\sigma+k_{1})x_{0}}}{2(k_{1}+\sigma)}e^{-k_{1}x}
+\frac{e^{-(k_{1}-\sigma)x_{0}}}{2(k_{1}-\sigma)}e^{k_{1}x}
\right].\label{e:omega}
\end{eqnarray}
Finally using the transformation $\rm t=e^x$, with $\rm x=\ln|t|$,
then the equation (\ref{e:trans-phi}) became
\begin{eqnarray}
\rm \nonumber \phi(t)=C_{1}\cosh(k_{1}\ln|t|)t^{-\frac{1}{2}a_{1}}+C_{2}\sinh(k_{1}\ln|t|)t^{-\frac{1}{2}a_{1}}\\
\rm \nonumber +\frac{d}{k_{1}}\left[\frac{k_{1}}{\sigma^2-k_{1}^2}
t^{\sigma-\frac{1}{2}a_{1}}+\frac{e^{(\sigma+k_{1})x_{0}}}{2(k_{1}+\sigma)}t^{-(k_{1}+\frac{1}{2}a_{1})}\right]\\
\rm+\frac{d}{k_{1}}\left[\frac{e^{-(k_{1}-\sigma)x_{0}}}{2(k_{1}-\sigma)}t^{k_{1}-\frac{1}{2}a_{1}}\right].\label{e:solution-phi}
\end{eqnarray}
Equation (\ref{e:solution-phi}) represent the solution to scalar field in the Radiation time and depend of the constants that appear there. Where $a_{1}=\frac{5-n}{n}$, $\sigma=\frac{3}{2}-\frac{3}{2n}$, $d=\frac{16}{3}\pi \frac{M_{1/3}}{(nD)^{4/n}}$, $k_{1}=\frac{1-n}{2n}$, note that all constants depend on $n$ and $0<n<1$.\\
\\Figure (\ref{infla-rad}) shows the shape of the scalar field $\phi$ in the inflation and radiation epoch, in the inflation time $\phi$ is growing, but \emph{in the transition between inflation and radiation epoch}, the scalar field slows its expansion and then decrease in the radiation time, perhaps due to the presence of the first particles. In the transition we must consider that the constant $M_{\gamma}$ must change for each stage of the universe.
\\

\textbf{Case III}\\
Stiff matter ($\gamma=1$), then the equation (\ref{e:escala3}) is written as
\begin{equation}\label{e:escala4}
2\frac{\ddot{A}}{A}+4\left(\frac{\dot{A}}{A}\right)^2+5\frac{\dot{A}}{A}\frac{\dot{\phi}}{\phi}
+\frac{\ddot{\phi}}{\phi}=0,
\end{equation}
substituting equation (\ref{e:escala2}) in (\ref{e:escala4}), yielding
\begin{eqnarray}
\nonumber \rm t^2\ddot{\phi}+\frac{5}{n}t\dot{\phi}+\left[\frac{6-2n}{n^2}\right]\phi=0, \\
\rm t^2\ddot{\phi}+at\dot{\phi}+b\phi=0, \label{e:casoI1}
\end{eqnarray}
with $a=\frac{5}{n}$, $b=\frac{6-2n}{n^2}$, with the following solution
\begin{equation}
\phi(t)=|t|^{\frac{1-a}{2}}(c_{1}|t|^{\mu}+c_{2}|t|^{-\mu}), \qquad (1-a)^{2}>4b,
\end{equation}\label{e:CasoI2}
where $c_{1}$, $c_{2}$ are constants and $\mu=\frac{n-1}{2n}$.\\
\\Then the solution to equation (\ref{e:casoI1}) is
\begin{equation}
\rm \phi(t)=c_{1}|t|^{1-\frac{3}{n}}+c_{2}|t|^{-\frac{2}{n}}.
\end{equation}
Figure (\ref{Scalar}) shows that the contribution of the scalar field at the beginning was significantly, but decreasing exponentially along the time, also we note that $\phi$ decreases more quickly than in the radiation time.
\begin{figure} [h]
\begin {center}
\includegraphics[width=8.5cm]{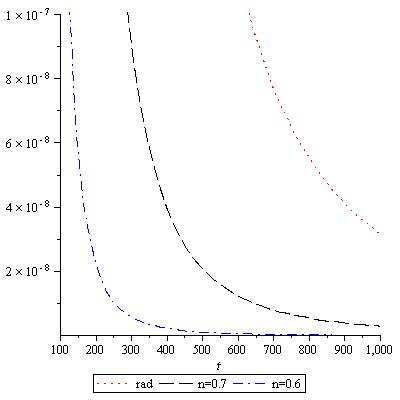}
\caption{The black and blue line represents the behavior of the scalar field in the stiff matter epoch. The red line represents the solution to the radiation epoch obtained
in the previous section, we note that in the transition between radiation and stiff matter, the scalar field decreases more quickly than predicted radiation curve. For very large times
 $\phi \rightarrow 0$. \label{Scalar}
}\label{Scalar}
\end {center}
\end{figure}
\\Finally, the physical quantities take the form:
\begin{equation}
\rm P=\rho=M_{1}(Dnt)^{\frac{-6}{n}}(c_{1}t^{1-\frac{3}{n}}+c_{2}t^{-\frac{2}{n}})^{\beta}. \label{e:scalar-1}
\end{equation}
where $\beta=\frac{\lambda_{0}(3+2\omega)}{2}-1$.\\

\begin{figure} [h]
\begin {center}
\includegraphics[width=8cm]{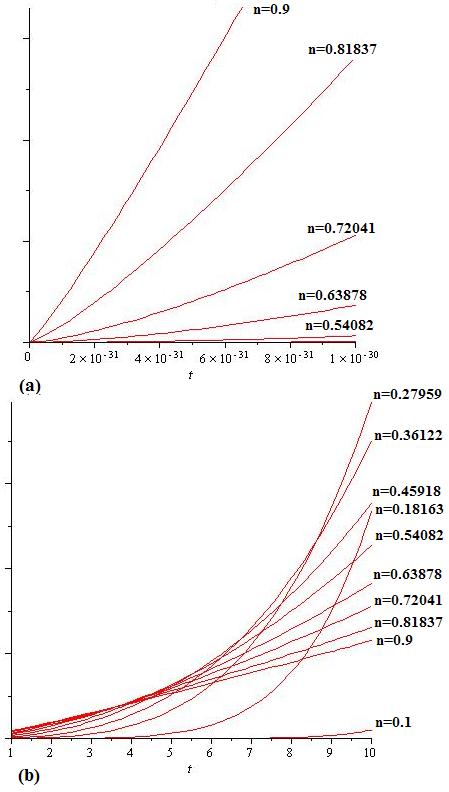}
\caption{(a) Behavior of the scalar factor in the inflation and radiation time, we note that for $0.6<n<1$ is when the universe has a rapid expansion. (b) after radiation time the shape of the scalar field is modified so the expansion scenary today correspond to $0.3<n<1$. $D$ is the slope of the curve, but this constant have
different values for each age of the universe. Berman \cite{Berman}
calculate this value for dust matter ($\gamma=0$) with $D\cong 5.0
\times10^{28}\frac{cm^2}{s}$.}\label{S-F}
\end {center}
\end{figure}

On the other hand, remembering that our results are based on the
Berman's law \cite{Berman}, we obtain an expression for the average
scalar factor (equations \ref{e:escala1}-\ref{e:escala2}), taking
the equation (\ref{e:escala2})(case $n\neq0$) to solve the
corresponding set of equations. Figure (\ref{S-F}) shows the
behavior of scalar factor to different values of $n$ ($0<n<1$). Note
that when $n$ increases, the scalar factor grows rapidly, i.e. the
slope of the curve is large, but for certain values of $n$, the
slope of the curve approaches a plane curve. Remembering $D$ is a
constant and in this case has the role of slope of the curve. With
the observational data in your epoch, Berman \cite{Berman}
calculated this value for dust matter, yielding to the value $D\cong
5.0 \times10^{28}\frac{cm^2}{s}$. Considering the shape of the scale
factor for some values in the n parameter, we say that the expansion
scenary today corresponds for a special value to n, in the branch
$\rm [0.3, 1)$ figure (\ref{S-F}(b)). These
intervals can be modified when one calculate the value to the
constant D considering the observational data today. This
calculation will be part of forthcoming paper.
\\

\section{Lagrangian and Hamiltonian densities in SCC}
In the previous section using the Berman's law, were solved the Einstein field equations.
Now we will use the classical approach to find solutions to $(\rho,A,\phi)$.
The corresponding Lagrangian density using (\ref{metric}) into (\ref{L1})
\begin{equation}
\rm{\cal L}= 6\frac{A \phi {\dot A}^2 }{N}+ 6\frac{A^2 \dot A \dot \phi}{N} -\frac{\omega A^3{\dot \phi}^2}{N \phi}+16\pi A^3 \rho N,
\end{equation}
the momenta are $(\Pi_{q}=\frac{\partial \cal L}{\partial\dot{q}})$
\begin{eqnarray}
&&\rm \Pi_A=\rm 12 \frac{A \phi \dot A}{N}+6\frac{A^2 \dot \phi}{N}, \qquad \Pi_\phi=\rm 6\frac{A^2 \dot A}{N}- 2\frac{\omega A^3 \dot \phi}{N \phi},\nonumber\\
&&\rm \dot A = \frac{N}{6(3+2\omega) \phi A^2}
\left( 3\phi \Pi_\phi+\omega A \Pi_A\right), \label{pia}\\
&&\rm \dot \phi = \frac{N}{2(3+2\omega) A^3}\left(A \Pi_A - 2 \phi \Pi_\phi \right), \label{piphi}
\end{eqnarray}
when we write the canonical Lagrangian density $\rm {\cal L}_{canonical}=\Pi_q \dot q- N{\cal H}$,
we obtain the corresponding Hamiltonian density as
\begin{eqnarray}
&&\rm {\cal H}= \frac{A^{-3}}{12a_0 \phi}[-6\phi^2 \Pi_\phi^2 +
\omega A^2 \Pi_A^2 +6A \phi \Pi_A \Pi_\phi \nonumber \\ &&
 -192 \pi a_0 A^6 \phi \rho], \qquad a_0=(3+2\omega), \qquad  \label{hami1}
\end{eqnarray}
\subsection{Classical scheme: Hamilton-Jacobi equation}
Using the gauge $\rm N= 12a_0 \phi A^3$, and the transformation in
the momenta $\rm \Pi_q=\frac{\partial S}{\partial q}$, where S is
known as the superpotential function, with this, the Hamiltonian
density is written as (we include
 the equation (\ref{rho}) in this
last equation)
\begin{eqnarray} &&\rm
-6\phi^2 \left(\frac{\partial S}{\partial \phi}\right)^2 + \omega A^2 \left(\frac{\partial S}{\partial A}\right)^2
 +6A \phi \left(\frac{\partial S}{\partial A}\right) \left(\frac{\partial S}{\partial \phi}\right) \nonumber
\\ &&\rm - 192\pi a_0 M_\gamma A^{-3(\gamma-1)} \phi^{(\beta+1)}=0, \label{hamilton-jacobi}
\end{eqnarray}
This is the differential equation in the Hamilton-Jacobi theory,
equation (\ref{hamilton-jacobi}), which can be solved for general
barotropic fluid using the method of Lagrange-Charpit
\cite{Elsgoltz,delgado,lopez}, solutions that will be presented
elsewhere. So in the follow, for simplicity we shall use the stiff
matter case, $\gamma=1$ (so
$\beta+1=\frac{\lambda_{0}(3+2\omega)}{2}$). Using $\rm
S(A,\phi)=S_1(A) +S_2(\phi)$, we have
\begin{eqnarray} &&\rm
-6\phi^2 \left(\frac{d S_2}{d \phi}\right)^2 + \omega A^2 \left(\frac{d S_1}{d A}\right)^2
 +6A \phi \left(\frac{d S_1}{d A}\right) \left(\frac{d S_2}{d \phi}\right) \nonumber
\\&&\rm -192 \pi a_0 M_1  \phi^{(\beta+1)}   =0, \label{hamilton-jacobi1}
\end{eqnarray}
identifiying $\rm A \frac{dS_1}{dA}=c_{1}=constant$, then (\ref{hamilton-jacobi1}) can be solve as
\begin{equation}
\rm \frac{dS_2}{d\phi}=\frac{c_{1}}{2\phi} \mp
\frac{1}{6\phi}\sqrt{3c_{1}^2a_{0}-1152\pi a_0 M_1 \phi^{(\beta+1)}},
\label{phii}
\end{equation}
using (\ref{pia}) and (\ref{piphi}) we find that $c_{1}a_{0}=\frac{\dot
A}{A}+\frac{1}{2} \frac{\dot \phi}{\phi}$, then (\ref{phii}) can be
determined as
\begin{equation}
\rm \phi(t)=\phi_0 \, sech^{\frac{2}{\beta+1}}\left(-c_{0}
\sqrt{3a_0} t \right),
\end{equation}
so, the solution for the scale factor A, become
\begin{equation}
\rm A(t)=A_{0}\cosh^{\frac{1}{\beta+1}} \left(-c_{0}\sqrt{3a_{0}} t
\right) \exp\left[\frac{c_{0}a_0 t}{\beta+1}\right].
\end{equation}
\\

\section{Conclusions}
In this paper we have investigated FRW cosmological model of the universe in the framework of Barber's second self-creation
 theory, we obtain a solution for three cases: inflation ($\gamma=-1$), radiation ($\gamma=\frac{1}{3}$) and stiff matter ($\gamma=1$),
 in the first case the scalar field was an increasing function and depends on the constant $M_{-1}$,
 we note that in the transition between inflation and radiation
 epoch, the scalar field stop its expansion and then decreases. This behavior is due to the constant $M_{\gamma}$
  that appear in the solution must change for each stage of the universe and we must also consider the presence of
  the first particles in the radiation time, in the stiff matter time the scalar field decrease exponentially and for
  very long time $\phi \rightarrow 0$. Such that currently, the scalar field contribution is very small.\\
 In addition we found a new parameter $\lambda_{0}\neq\lambda=\frac{2}{3+2\omega}$ more general than barber's parameter,
 but when $\lambda_{0}=\lambda$ the solution of density (\ref{rho}) is the same that in general relativity, so there
  is no contribution of the scalar field.
 Our results are preliminary with this and will depend on how to adjust the constant with current observations.
 On the other hand, the classical solution found under the Hamilton-Jacobi approach (stiff matter) have an
 structure more general. Moreover, the feeling is that in some approximation, this class of solution could be
  tied with the  Berman's law.

\acknowledgments{This work was supported in part by  DAIP
(2011-2012), Promep  UGTO-CA-3 and CONACyT 167335 and 179881 grants.
JMR was supported by Promep grant ITESJOCO-001. Many calculations
where done by Symbolic Program REDUCE 3.8. This work is part of the
collaboration within the Advanced Institute of Cosmology  and Red
PROMEP: Gravitation and Mathematical Physics under project {\it
Quantum aspects of gravity in cosmological models, phenomenology and
geometry of space-time}.}
\\

 \end{document}